# The Influence of Cultural Distance on Settlement Intention of Floating Population in China

Dan Qin

ucfndqi@ucl.ac.uk

## Abstract

Based on a nationwide labour-force survey data, this paper investigates the influence of cultural variance on migrants' settlement intention in China. By using dialectal distance as a proxy for cultural distance, we find strong evidence for the negative effects of cultural distance on migrants' settlement intention. By further investigation into sub-samples separated by gender, generation and higher education experience, we find that the influence is less effective for younger migrants and higher-educated migrants, which indicates that the impact of cultural barrier may gradually diminish with the integration of society and promotion of education.

## 1. Introduction

The increase in the number of migrants has led to a growing number of floating populations in China. Although we would generally assume that most migrants move in order to settle down, according to the China Labour-force Dynamic Survey (2014), nearly half of the floating population do not intend for permanent stay in the destination city.

Numerous discussions have been proposed on the determinants of migrants' settlement intentions and two theoretical approaches predominate in the debate. The first one, built around the neo-classical economic theory, emphasizes the importance of economic incentives to migrants' decision-making. The other underlines the impact of socio-cultural conditions and believes that an individual's social attachment towards the destination place has a significant effect on one's tendency to settle.

In the discussion of socio-cultural factors, previous studies mainly concentrate on the pull effect of migrants' social connections with the city, whereas the push effect of cultural variance between a migrant's origin and destination place is rarely discussed. In fact, the negative influence of cultural difference is already apparent when migrants make decisions on the choice of destination places. Evidence has been found in both international migration (Belot and Ederveen, 2012) as well as migration within China (Li and Meng, 2014; Liu, Xu and Xiao, 2015). The theoretical account behind the finding could be that people in culturally homogeneous groups tend to have less difficulty in developing their level of trust than people from culturally heterogeneous groups, and thus people are more willing to flow to areas similar to their birthplaces.

Therefore, in this paper, we would like to investigate the influence of cultural distance on migrants' settlement decisions. If cultural distance indeed contributes to one's communication cost, we may be able to see a negative impact on migrants' settlement intention. Based on a national labour-force survey data, we use logistic regression to estimate the migration decision model. Furthermore, we extend our analysis by exploring the impact in multiple subpopulations, separated by gender, generation and higher-education experience.

In what follows, we will give an overview of the related literature in [Section 2](). In [Section 3]() we discuss the data and methods used in our analysis, including cultural distance measures and other quantitative methods. Results and discussions are presented in [Section 4]() and conclusion is given in the final section.

## 2. Literature Review

### 2.1 Determinants of settlement intention

As is mentioned, there are two competing theories on the determinants of migrants' settlement intentions. The first one describes the migrants' decision-making process as a rational behaviour and choices are made in order to maximise their economic profits (Constant and Massey, 2003). Therefore, they believe that migrants' willingness to settle is positively correlated with economic incentives. More specifically, education, which is an important factor of human capital formation, significantly contribute to one's economic capabilities (Blundell et al., 2005; Lange and Topel, 2006). Proficiency of local dialect, on the other hand, helps to minimise communication costs and promote the acquisition of social capital in local society (Mesch, 2003). The status in the labour market, usually represented by income level and stability of occupation, implies higher level of economic achievements. Overall, the more likely a migrant is able to achieve economic success in the destination place, the more willing he is to settle down in the city.

Another theory puts emphasis on migrants' cultural and social attachment towards the city and believes that migrants with a stronger social bond with the destination are more intended for long-term settlement. The number of co-living family members, for example, significantly strengthen the migrants' will to settle in the city (Chen and Liu, 2016). Similarly, the size of one's local social networks also has a positive impact on migrant's settlement intention, whereas social connections with the origin place decrease the tendency to settle (Haug, 2008). In addition, the number of years the migrants have stayed in the destination city contributes to the accumulation of human and social capitals and therefore intensifies both economic and socio-cultural impacts.

Moreover, the institution barrier caused by household registration system (also known as Hukou system) is also an influential factor in studies of migration settlement in China. First established in 1958, the two-fold household registration system, which is highly linked with social welfare programs and assigns benefits based on rural and urban residency status, restricts migrants' equal access to social opportunities. The negative influence of the system's constraints on migrants' settlement intention has been proved by numerous literature (Wang and Fan, 2012; Tang and Feng, 2015; Huang and Cheng, 2014; Zhu and Chen, 2009). Although some literature argues that its impact has declined substantially because of the market reforms and policy changes (Zhan, 2011), we still include the factor in our analysis considering the time span of our data.

### 2.2 Dialectal distance as a proxy of cultural distance

There are several measuring methods of cultural distance that have been frequently adopted in previous studies.

The first one is proposed by Hofstede (1984) who used five cultural dimensions to evaluate the cultural variance between countries based on an IBM employee survey from across 53 countries. Subsequent researchers developed on his dimension model and their findings are widely used in studies of relationship between culture and economics (Belot and Ederveen, 2012). Although being probably the most widely applied cultural distance measure, no available data conducted under the same criteria can be found in China, therefore it is not applicable to our analysis.

Next we find another popular measure of cultural distance, which is genetic distance. Spolaore and Wacziarg (2009) first developed the theoretical account behind the correlation between genetic distance and cultural variance, and described genetic distance as a summary measure of long-term divergence in the traits across populations. However, it is later pointed out in some literature that genetic distance is not an appropriate variable to capture cultural traits in China (Zhao and Lin, 2017). This is because genetic distance mainly describes the inheritance of culture, but fails to reflect the horizontal transmission of culture, whereas the formation of the Chinese nation involves frequent integration with neighbouring minorities.

According to our aim of use, we alternatively choose linguistic distance as a proxy variable to describe cultural distance in China. On the one hand, language is able to capture the variance in both vertical and horizontal transmission of culture. On the other hand, the Chinese languages manifest a huge dialectal variety and the distribution of dialects are generally associated with the distribution of regional cultures. Previous studies have also adopted this measure in economic studies in China (Gong, Chow and Ahlstrom, 2011; Lin and Zhao, 2017; Gao and Long, 2016; Herrmann-Pillath, Libman and Yu, 2014). In this paper, we use the population weighting method proposed by Liu, Xu and Xiao (2015) to calculate the dialectal distance. Details will be explained in Section 3.2.

## 3. Data and methodology

3.1 Models and variables

Binary logistic regression is chosen to estimate the effect of cultural distance on migrants' settlement intention. We first estimate the model with the full sample, and then further investigate the variance of the effect in several sub-populations.

The first grouping is divided by gender, for males and females may vary in the ways of integration in a new environment and we are curious whether cultural difference has equal impacts for both genders. The next one is separated by generation. New generation of migrants who were born after the 1980s, are considered to have different concerns on settlement from the first generation (Liu, Li and Breitung, 2012; Cheng, Wang and Smyth, 2014). Chen and Liu (2016)'s study suggests that settlement decision of the new generation is more driven by economic incentives compared with the first generation. Furthermore, we would also like to investigate the influence of higher education experience on the impact of cultural variance.

As for the data, we retrieve the labour-force data from the China Labour-force Dynamic Survey (CLDS, 2014), which is a nationwide survey of samples from 29 out of 34 provinces and covers issues at individual, family and community levels. For the dependent variable, we recode answers for the question "how do you feel like settling in the current city" to binary values. That is, "impossible", "unlikely" and "not sure" are coded as 0 (i.e., not willing to settle), whereas "possible" and "highly possible" are coded as 1 (i.e., willing to settle). For economic incentives, we introduce the migrant's education level, proficiency of local dialect, job satisfaction and length of stay in the destination as control variables. The socio-cultural conditions are described by the number of co-living families, count of local friends and frequency of interaction with neighbours. We also control for the institutional barrier, which is implied by hukou type. All the above variables can be found in CLDS (2014) dataset.

As for the independent variable, we find the dialect data from the Language Atlas of China (LAC, 2012) and the Chinese Dialect Dictionary (CDD, 1991). These two references allow us to get the dialect spoken in each county in China.

Apart from the determinants that are fully discussed in previous literature, we would also like to include regional factors to control for the potential regional impacts. GDP per capita and population of the origin and destination places are therefore extracted from the China City Statistical Yearbook (CCSY) and other regional statistical bulletins. As we find annual average of raw GDP values is difficult to interpret, we calculate the relative ratio of GDP per capita of each prefecture to the value of Shanghai for each year, and then average the ratios by year. And we do the same for the population.

County population is retrieved from the sixth China Population Census (2010) for the calculation of population weighting method. Furthermore, as geographic distance has a high correlation with dialectal distance, we also control for the variable and acquire the distance data from BaiduMap's public API.

3.2 Dialectal distance measures

Two measures of dialectal distance are most widely applied in the current literature. One is based on the branch distance of linguistic trees, the other is by calculating lexicostatistical distance.

By computing the proportion of resemblances between language groups, the lexicostatistical distance is more scientifically-founded than distances based on linguistic trees. Cheng (1997) employed similar approaches to quantify dialect affinity by calculating the correlation coefficients of phonological elements of 17 Chinese dialects. However, the dataset is not large enough to cover our data (only 9 samples are matched) and it is not applicable for our study.

Therefore, we use the branch distance of linguistic trees as a proxy for dialectal distance. Due to historical reasons, boundaries of prefecture divisions [1] break up the originally homogeneous culture areas and we usually find multiple dialects being used within the same prefecture. However, location data of migrants is only available at prefecture-level for privacy concerns, therefore we aggregate county-level dialectal data to prefecture level by using the population weighting method proposed by Liu, Xu and Xiao (2015).

We first extract county population data from the census and calculate the population proportion of counties in each prefecture. We then generate combinations of all counties and for each combination, we acquire the dialectal distance based on the following assignment rules:
1. If two counties belong to the same sub-dialect group, the distance is 0

2. If two counties belong to the same dialect group, but different sub-dialect groups, the distance is 1

3. If two counties belong to the same super dialect group, but different dialect groups, then the distance is 2

---

[1] Administrative divisions of China (Wiki): https://en.wikipedia.org/wiki/Administrative_divisions_of_China

4. If two counties belong to the same language branch, but different super dialect groups, the distance is 3
5. If two counties belong to different branches, the distance is then 4

Thus, we have the dialectal distances between each combination of counties. Next, we use the share of county population to calculate the dialectal distance between prefectures. Let us use $d(A,B)$ to denote the dialectal distance between prefecture $A$ and prefecture $B$. Then the population proportion of county $i$ in prefecture $A$ is denoted by $S_{Ai}$, and the population proportion of county $j$ in prefecture $B$ by $S_{Bj}$. With the dialectal distance between county $i$ and county $j$ denoted by $d_{ij}$, the dialectal distance between prefecture $A$ and prefecture $B$ is then:

$$d(A,B) = \sum_{i=1}^{I} \sum_{j=1}^{J} S_{Ai} * S_{Bj} * d_{ij} \quad (1)$$

The calculated value $d(A,B)$ can be interpreted as the expectation of dialectal distance between a random resident from prefecture $A$ and another random resident from prefecture $B$. Table 1 gives an example of the dialectal distance results. The non-integer values suggest that there are multiple dialects being used in the city and the within-prefecture distances indicate the dialect diversity in each city.

Table 1 Dialectal distance of prefectures

|           | Beijing | Guangzhou | Shanghai | Nanjing | Nanning |
|-----------|---------|-----------|----------|---------|---------|
| Beijing   | 0.083   | 3         | 3        | 2.1048  | 2.7628  |
| Guangzhou | 3       | 0         | 2        | 2.8952  | 1.4743  |
| Shanghai  | 3       | 2         | 0        | 2.7905  | 2.2372  |
| Nanjing   | 2.1048  | 2.8952    | 2.7905   | 0.5627  | 2.7078  |
| Nanning   | 2.7628  | 1.4743    | 2.2372   | 2.7078  | 1.0855  |

## 4. Estimation results

### 4.1 Full sample model

Table 2(col.1) presents the estimation results of economic incentives for the full sample, including controls for regional and institutional variables.

Let us first comment on the effects of regional and institutional variables. We find that GDP per capita of the destination has a significant negative correlation with migrants' settlement intention, whereas GDP per capita of the origin is positively correlated with the dependent variable. This implies that GDP per capita actually has a push effect on migrants' settlement decision-making. The story could be that high GDP per capita of the city is generally associated with high pressure of competition and high price level, especially high prices of housing, which impede migrants from settling. The odds ratio presented in Table 2(col.1-1) helps us to understand the magnitude of the effects. With one unit increase in the relative ratio of GDP per capita of destination city, the odds of settlement intention decreases by 55.6%. Whereas one unit increase of the ratio of origin increases settlement intention by 77.2%. Migrants' hukou type also has a significant positive effect on settlement intention and the odds of urban migrants' tendency for settlement is 2.65 times larger than the odds for rural migrants.

Turning to the role of economic incentives, we find that the increase in migrants' education experience and job satisfaction, as well as the improvement in local dialect proficiency all have significant positive effects on settlement intention. This is an indication that migrants with better opportunities of economic achievements are more likely to settle down in the city.

Table 2 Determinants of migrant's settlement intention

|  | (1) Coeff. (Std.) | (1-1) OR (95% CI) | (2) Coeff. (Std.) | (2-1) OR (95% CI) |
| --- | --- | --- | --- | --- |
| Gender | 0.161(0.1401) | 1.175(0.893-1.547) | 0.1536(0.1427) | 1.166(0.882-1.544) |
| Age | 0.0127(0.0073) | 1.013(0.998-1.028) | 0.0121(0.0073) | 1.012(0.998-1.027) |
| Hukou type | 0.9761(0.1843)*** | 2.654(1.854-3.82) | 0.9839(0.1861)*** | 2.675(1.862-3.864) |
| GDP/cap ori. | 0.5722(0.238)* | 1.772(1.101-2.813) | 0.5552(0.2389)* | 1.742(1.081-2.773) |
| GDP/cap dest. | -0.8109(0.1162)*** | 0.444(0.35-0.552) | -0.7871(0.1164)*** | 0.455(0.358-0.566) |
| Pop. origin | 0.01(0.2415) | 1.01(0.62-1.604) | -0.0278(0.244) | 0.973(0.594-1.553) |
| Pop. dest. | -0.0331(0.3025) | 0.967(0.535-1.753) | 0.0544(0.3048) | 1.056(0.581-1.924) |
| Education level | 0.4997(0.1187)*** | 1.648(1.309-2.084) | 0.5431(0.1207)*** | 1.721(1.361-2.186) |
| Dialect profi. | 0.4381(0.0494)*** | 1.55(1.409-1.71) | 0.4296(0.0501)*** | 1.537(1.395-1.698) |
| Len. of stay (y) | 0.0115(0.0057)* | 1.012(1-1.023) | 0.0108(0.0057) | 1.011(1-1.022) |
| Job satisfaction | 0.2472(0.089)** | 1.28(1.077-1.527) | 0.2451(0.0905)** | 1.278(1.071-1.528) |
| Num. of co-living family |  |  | 0.1523(0.0492)** | 1.164(1.058-1.283) |
| Num. of local friends |  |  | 0.0133(0.006)* | 1.013(1.001-1.025) |
| Neighbour relationship |  |  | 0.085(0.07) | 1.089(0.949-1.249) |
| Resid. Df | 1272 |  | 1269 |  |
| AIC | 1311.9 |  | 1300 |  |

*p = 0.05; **p = 0.01; ***p = 0.001

By introducing the socio-cultural factors into the model (Table 2(col.2)), we find a significant and positive correlation between settlement intention and local social networks. For each additional member living together in the family, the odds of migrants' tendency to settle increases by 16.4%. Whereas by knowing one more person in the local society increases the odds by 1.3%. Migrants' connections with neighbours, on the other hand, do not have significant impacts. We also notice length of stay in the destination no long has significant effects after controlling for the socio-cultural factors.

Finally, we turn to the role of cultural distance between places (Table 3(col.3)). We find that for each one unit increase in the dialectal distance, which in this case indicates that the divergence of two dialects occurs at a higher level in the linguistic trees, results in a decrease by 22.9% in the odds of settlement intention. The negative effect of dialectal distance remains significant even when controlling for geographic distance (Table 3(col.4)).

Furthermore, we compare the fitness of models by calculating AIC (Akaike information criterion) values. By introducing socio-cultural factors, the AIC of the model decreases and it continues to decrease as we include the dialectal distance, which implies that the new variables do help improve the fitness of the model. Binned residual plots are also created to check the residuals. From Fig.1, we can see that most of the points locate within the confidence bands (grey lines) and no obvious pattern can be observed in the plots, which indicates we have adequate models and well-behaved data. As the effect of geographic distance is not significant and it does not help improve the model fit, we decide to exclude the variable in later analysis.

Table 3 Determinants of migrant's settlement intention - the role of culture

|  | (3) Coeff. (Std.) | (3-1) OR (95% CI) | (4) Coeff. (Std.) | (4-1) OR (95% CI) |
|---|---|---|---|---|
| Gender | 0.1425(0.1434) | 1.153(0.871-1.529) | 0.1446(0.1435) | 1.156(0.873-1.532) |
| Age | 0.0137(0.0074) | 1.014(0.999-1.029) | 0.0134(0.0074) | 1.014(0.999-1.029) |
| Hukou type | 0.996(0.1871)*** | 2.707(1.881-3.919) | 0.9886(0.1874)*** | 2.687(1.866-3.893) |
| GDP/cap ori. | 0.2549(0.2629) | 1.29(0.761-2.144) | 0.2483(0.2628) | 1.282(0.756-2.13) |
| GDP/cap dest. | -0.7055(0.1157)*** | 0.494(0.389-0.613) | -0.6963(0.1162)*** | 0.498(0.392-0.619) |
| Pop. origin | 0.0161(0.248) | 1.016(0.616-1.635) | -0.0199(0.2544) | 0.98(0.587-1.596) |
| Pop. dest. | 0.0731(0.3084) | 1.076(0.589-1.976) | 0.1184(0.3163) | 1.126(0.607-2.1) |
| Education level | 0.5397(0.1212)*** | 1.716(1.356-2.181) | 0.5415(0.1212)*** | 1.719(1.358-2.185) |
| Dialect profi. | 0.3626(0.0547)*** | 1.437(1.292-1.602) | 0.3676(0.0553)*** | 1.444(1.297-1.612) |
| Len. of stay (y) | 0.0051(0.006) | 1.005(0.993-1.017) | 0.0055(0.0061) | 1.005(0.994-1.018) |
| Job satisfaction | 0.2591(0.0911)** | 1.296(1.085-1.552) | 0.2578(0.0912)** | 1.294(1.084-1.55) |
| Num. of co-living family | 0.1528(0.0494)** | 1.165(1.058-1.284) | 0.1512(0.0495)** | 1.163(1.056-1.282) |
| Num. of local friends | 0.0139(0.006)* | 1.014(1.002-1.026) | 0.0139(0.006)* | 1.014(1.002-1.026) |
| Neighbour relationship | 0.0905(0.0704) | 1.095(0.954-1.257) | 0.0908(0.0704) | 1.095(0.954-1.258) |
| Dialectal distance | -0.2605(0.0901)** | 0.771(0.645-0.919) | -0.2977(0.1064)** | 0.742(0.602-0.914) |
| Geographic distance (km) |  |  | 1.050e-07 (1.587e-07) | 1(1-1) |
| Resid. Df | 1268 |  | 1267 |  |
| AIC | 1293.5 |  | 1295.1 |  |

*$p = 0.05$; **$p = 0.01$; ***$p = 0.001$

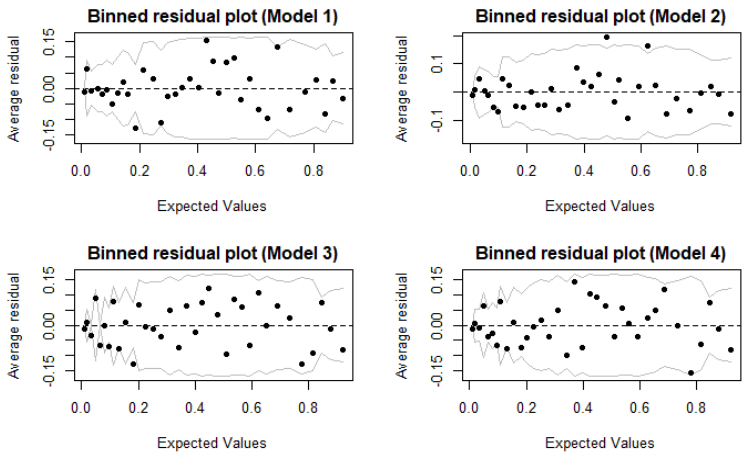

Fig.1 Binned residual plots of four models

## 4.2 Sub-sample models

Turning to the sub-sample models, we first look at the estimation results for different gender groups (Table 4). We find that the positive influence of hukou type remains significant for both groups. As for the economic incentives, GDP per capita of the destination, education level and proficiency of local dialect still have significant impacts for both male and female migrants. Job satisfaction, interestingly, has a significant positive effect on males, but the effect is not of equal importance for the females, which indicates that job satisfaction is less effective in determining female migrants' settlement decisions. The influence of socio-cultural conditions is similar for both genders and contrary to findings of the full sample model, the effect of local social network is not significant in either groups.

Dialectal distance has a significant negative effect on both genders' settlement intention. And for each unit increase in the dialectal distance, the odds of settlement intention decreases by 23.3% for the females and 25.8% for the males. To further compare the gender variance in the coefficients of dialectal distance, we introduce an interaction term of gender and dialectal distance, denoted by *Gender\*Dia_dist* ([Table 4](col.7)). We find that the p-value of *Gender\*Dia_dist* is beyond our significance level, which implies that there is no significant difference in the influence of dialectal distance between males and females.

Table 4 Determinants of migrant's settlement intention - gender subgroups

|  | (5)-female Coeff. (Std.) | (5-1) OR (95% CI) | (6)-male Coeff. (Std.) | (6-1) OR (95% CI) | (7) |
|---|---|---|---|---|---|
| Age | 0.0145(0.0105) | 1.015(0.994-1.036) | 0.0137(0.0105) | 1.014(0.993-1.035) | 0.0137(0.0074) |
| Hukou type | 0.7244(0.2713)** | 2.063(1.215-3.526) | 1.2587(0.2654)*** | 3.521(2.108-5.978) | 0.9942(0.1872)*** |
| GDP/cap ori. | 0.3687(0.3556) | 1.446(0.71-2.915) | 0.0724(0.4063) | 1.075(0.467-2.311) | 0.2608(0.2628) |
| GDP/cap dest. | -0.6861(0.1675)*** | 0.504(0.353-0.685) | -0.7227(0.1626)*** | 0.485(0.344-0.654) | -0.7069(0.1158)*** |
| Pop. origin | -0.1275(0.3905) | 0.88(0.391-1.826) | 0.1031(0.3327) | 1.109(0.568-2.112) | 0.0206(0.2482) |
| Pop. dest. | 0.26(0.4528) | 1.297(0.536-3.18) | -0.1111(0.437) | 0.895(0.381-2.129) | 0.0689(0.3088) |
| Education level | 0.3567(0.157)* | 1.429(1.054-1.953) | 0.8995(0.1973)*** | 2.458(1.679-3.643) | 0.5422(0.1213)*** |
| Dialect profi. | 0.3448(0.0746)*** | 1.412(1.222-1.638) | 0.3919(0.0823)*** | 1.48(1.263-1.745) | 0.3637(0.0548)*** |
| Len. of stay (y) | -0.0006(0.0084) | 0.999(0.983-1.016) | 0.0123(0.0089) | 1.012(0.995-1.03) | 0.0051(0.006) |
| Job satisfaction | 0.2364(0.1312) | 1.267(0.982-1.644) | 0.3013(0.13)* | 1.352(1.05-1.75) | 0.2588(0.0912)** |
| Num. of co-living family | 0.171(0.0683)* | 1.186(1.039-1.358) | 0.1477(0.0744)* | 1.159(1.002-1.342) | 0.1527(0.0494)** |
| Num. of local friends | 0.0178(0.011) | 1.018(0.996-1.041) | 0.0121(0.0075) | 1.012(0.997-1.027) | 0.014(0.0061)* |
| Neighbour relationship | 0.1573(0.1002) | 1.17(0.962-1.426) | 0.0161(0.1036) | 1.016(0.829-1.245) | 0.0875(0.0706) |
| Dialectal distance | -0.2659(0.1266)* | 0.767(0.597-0.982) | -0.2987(0.1308)* | 0.742(0.572-0.957) | -0.295(0.1114)** |
| Gender |  |  |  |  | 0.0599(0.2114) |
| Gender * Dia_dist |  |  |  |  | 0.0705(0.1325) |
| Resid. Df | 614 |  | 640 |  | 1267 |

*p = 0.05; **p = 0.01; ***p = 0.001

Next, we turn to the comparison between generations. As is indicated in Table 5(col.8&col.9), the effect of institutional barrier is no longer significant for the new generation. However, for the older generation, it is still an effective factor and the odds of settlement intention is about 4.3 times higher for urban migrants compared with rural migrants. While education level and job satisfaction are influential in terms of settlement decisions for younger migrants, the older generation is not equally affected. The number of local friends has a significant positive effect on older migrants, whereas it is not significant with the younger generation. This is an indication that the new generation is more driven by economic incentives, while the older generation is more dependent on social networks in the local society, which is in line with previous studies (Chen and Liu, 2016).

The influence of dialectal distance also differs among two generations. Although the effect is negative for both groups, it is only significant for the older migrants, which implies that the new generation is more tolerant of cultural variance between areas.

Table 5 Determinants of migrant's settlement intention - generation and education subgroups

|  | (8)-young Coeff. (Std.) | (9)-old Coeff. (Std.) | (10)-higher edu. Coeff. (Std.) | (11)-not higher edu. Coeff. (Std.) |
|---|---|---|---|---|
| Gender | 0.173(0.2123) | 0.0863(0.2033) | -0.2571(0.33) | 0.196(0.1598) |
| Age | 0.0316(0.0259) | 0.0222(0.0128) | 0.0022(0.0152) | 0.0128(0.0081) |
| Hukou type | 0.4223(0.2714) | 1.4578(0.2731)*** | 0.6615(0.3376) | 1.078(0.2324)*** |
| GDP/cap ori. | 0.0769(0.3582) | 0.4672(0.438) | -0.4216(0.6453) | 0.4229(0.2917) |
| GDP/cap dest. | -0.6496(0.1604)*** | -0.7578(0.1735)*** | -0.4409(0.2699) | -0.7783(0.1342)*** |
| Pop. origin | -0.0708(0.3531) | 0.2084(0.3591) | -0.3736(0.601) | 0.3035(0.2739) |
| Pop. dest. | 0.804(0.4428) | -0.7156(0.4449) | 1.7505(0.6478)** | -0.6165(0.3659) |
| Education level | 0.9661(0.21)*** | 0.2846(0.1556) |  |  |
| Dialect profi. | 0.2693(0.0807)*** | 0.4602(0.0791)*** | 0.2631(0.1297)* | 0.4144(0.063)*** |
| Len. of stay (y) | 0.0209(0.0137) | -0.0011(0.0071) | -0.0072(0.016) | 0.006(0.0065) |
| Job satisfaction | 0.285(0.142)* | 0.2364(0.125) | 0.2242(0.2331) | 0.2621(0.1008)** |
| Num. of co-living family | 0.1571(0.0696)* | 0.1622(0.0775)* | 0.3184(0.1215)** | 0.1396(0.0572)* |
| Num. of local friends | 0.0081(0.0098) | 0.0187(0.0082)* | 0.0207(0.0185) | 0.0125(0.0067) |
| Neighbour relationship | 0.0919(0.1074) | 0.0835(0.0975) | -0.0104(0.178) | 0.129(0.079) |
| Dialectal distance | -0.2154(0.1324) | -0.294(0.1302)* | -0.3934(0.2017) | -0.2737(0.105)** |
| Resid. Df | 604 | 648 | 231 | 1023 |

*$p = 0.05$; **$p = 0.01$; ***$p = 0.001$*

Finally, we turn to the results of education subgroups (Table 5 (col.10&col.11)). We find that for the higher-educated migrants, the effect of GDP per capita of the destination is not significant anymore, whereas population of the destination becomes significantly influential. The story could be that cities where people tend to gather are generally places with vibrant economies and better job opportunities. Meanwhile, as education experience is highly valued in the labour market in China, higher-educated workers are usually more favoured in competitive cities where GDP per capita could be high, therefore the negative effect of GDP per capita on higher-educated migrants is not as influential as migrants without higher education.

As for cultural influence, we find that dialectal distance remains significantly correlated with settlement intention of not higher-educated migrants, while for higher-educated group the negative effect is not significant. This further supports our assumption that the higher-educated population is more open-minded and adaptive to cultural difference between areas.

### 4.3 Robustness analysis

As we transformed the dependent variable from ordinal to binary data, it may cause problems due to the loss of information. Therefore, we first replace the dependent variable with the original ordinal data and use an ordinal logistic model to estimate the results (Table 6 (col.12)). We find that when holding all other variables constant, we can still observe a negative and significant correlation between dialectal distance and settlement intention.

The next issue is to distinguish between language and culture. Since we adopt dialectal distance as a proxy for cultural distance, we need to differentiate whether the effect is caused by language or by culture. Apart from the proficiency of local dialect, we also introduce a variable to control for the migrant's Mandarin proficiency. From Table 6 (col.13), we can see that when controlling for language capabilities, the negative effect of dialectal distance is still significant, which suggests that the influence is not related to language issues, but more possibly to cultural variance.

Table 6 Determinants of migrant's settlement intention - robustness analysis

|  | (12) | (13) | (14) | (15) |
|---|---|---|---|---|
| Gender | 0.2721(0.1103)* | 0.1422(0.1434) | 0.028(0.1742) | 0.0483(0.1735) |
| Age | 0.009(0.0055) | 0.0125(0.0076) | 0.0163(0.0089) | 0.0167(0.0089) |
| Hukou type | 1.0517(0.1505)*** | 1.005(0.1875)*** | 0.9802(0.2079)*** | 1.0066(0.2076)*** |
| GDP/cap ori. | -0.0263(0.2032) | 0.2747(0.2634) | 0.3988(0.2976) | 0.4211(0.2911) |
| GDP/cap dest. | -0.3692(0.0632)*** | -0.7046(0.1155)*** | -0.8353(0.1625)*** | -0.6429(0.1277)*** |
| Pop. origin | 0.0527(0.177) | 0(0.2493) | 0.3386(0.2832) | 0.2585(0.2809) |
| Pop. dest. | 0.2784(0.2397) | 0.0785(0.3089) | -0.4927(0.4633) | 0.0799(0.3739) |
| Education level | 0.3864(0.093)*** | 0.5619(0.1252)*** | 0.8512(0.1564)*** | 0.8913(0.1562)*** |
| Dialect profi. | 0.3884(0.0413)*** | 0.3603(0.0548)*** | 0.3242(0.0645)*** | 0.324(0.0642)*** |
| Len. of stay (y) | 0.0062(0.0048) | 0.0052(0.006) | -0.0002(0.0076) | 0.0012(0.0076) |
| Job satisfaction | 0.2302(0.0704)** | 0.2613(0.0911)** | 0.2017(0.1108) | 0.2103(0.1106) |
| Num. of co-living family | 0.1714(0.038)*** | 0.1543(0.0494)** | 0.1415(0.0613)* | 0.1361(0.061)* |
| Num. of local friends | 0.0106(0.0048)* | 0.0138(0.006)* | 0.0112(0.0076) | 0.0112(0.0076) |
| Neighbour relationship | 0.0604(0.0548) | 0.0875(0.0706) | 0.1225(0.0904) | 0.1291(0.0901) |
| Dialectal distance | -0.2137(0.0707)** | -0.2531(0.0906)** | -0.1888(0.108) | -0.1599(0.1066) |
| Mandarin profi. |  | 0.0681(0.095) |  |  |
| Policy restriction |  |  | 1.0092(0.508)* |  |
| Resid. Df | 1268 | 1267 | 918 | 919 |

*p = 0.05; **p = 0.01; ***p = 0.001*

Finally, we would like to introduce another control variable into our model, which is the policy restrictions for household registration in different cities. The influence of policy restrictions is rarely discussed in previous literature, however, as household registration is highly linked with welfare entitlements such as education, welfare housing and health care, it could have an impact on migrants' settlement decisions.

We use the degree of settling difficulty from Liu (2016), who calculated the difficulty to obtain local household registration in 63 large and medium-sized cities in China by aggregating weighted scores according to their settlement policies. As is shown in Table 6 (col.14), we find that, contrary to our expectations, the policy restriction is positively correlated with settlement intention at a significant level. A possible explanation could be that popular destinations for migrants usually have to impose strict standards for settlement, and there may be some common features of these cities that our model fails to capture. Therefore, the effects of these unknown features are explained by the coefficient of policy restriction.

We also find that the effect of dialectal distance is not significant when controlling for policy restrictions. As the sample size has changed after including the new variable, we also estimate the model with the same data but exclude the variable this time (Table 6 (col.15)). We see that opposed to the results with the same predictors in Table 3 (col.3), the coefficient of dialectal distance remains insignificant, which implies that the distinction may be caused by the variance in sampling areas. Therefore, in general, we cannot deny the importance of cultural distance to migrants' settlement decisions. But the finding does suggest that there are other factors that are more important than cultural variance for settlement in large and medium-sized cities.

## 5. Conclusion

Based on data from a nationwide labour-force survey (CLDS, 2014), this paper provides unique evidence for the role of cultural variance in migrants' settlement decision-making process. By using dialectal distance as a proxy for cultural distance, we find a significant negative correlation between cultural distance and migrants' settlement intentions.

We also extend our analysis by investigating into several subpopulations, separated by gender, generation and higher education experience. We find that for both male and female migrants, dialectal distance is negative correlated with the tendency to settle and there is no significant difference between the two genders. For the older generation and for migrants without higher education experience, dialectal distance has a strong hindering effect during the settlement decision-making process. Whereas for the new generation and higher-educated migrants, the influence is not significant.

Finally, we check the robustness of our findings by estimating the model with original ordinal data for the dependent variable. We also distinguish between language and culture by introducing the migrant's Mandarin proficiency and we find that the result remains robust when controlling for language capability variables. Furthermore, we also control for the influence of policy restrictions. The introduction of the new variable leads to a bias in sampling areas and we find that the negative effect of cultural distance fades away when migrants flow to large and medium-sized cities.

Although we believe that linguistic distance is an appropriate indicator of cultural variance, our method for measuring linguistic distance is still coarse due to limitation of data, which limits our precision in the results.

In conclusion, our findings suggest that cultural distance plays an important role in migrants' settlement decisions in contemporary China. The effect, however, may gradually diminish with the promotion of education and the integration of society.

The Influence of Cultural Distance on Settlement Intention of Floating Population in China | Dan Qin